# ChatMyopia: An AI Agent for Pre-consultation Education in Primary Eye Care Settings

Yue Wu, Xiaolan Chen, Weiyi Zhang, Shunming Liu, Wing Man Rita Sum, Xinyuan Wu, Xianwen Shang, Chea-su Kee, Mingguang He, Danli Shi

**Affiliation**

1. School of Optometry, The Hong Kong Polytechnic University, Hong Kong, China
2. Department of Ophthalmology, Guangdong Eye Institute, Guangdong Provincial People's Hospital, Southern Medical University, Guangzhou, China
3. Research Centre for SHARP Vision (RCSV), The Hong Kong Polytechnic University, Hong Kong, China
4. Centre for Eye and Vision Research (CEVR), 17W Hong Kong Science Park, Hong Kong, China

#Contributed equally

**\*Correspondence**

Dr. Danli Shi, MD, PhD. The Hong Kong Polytechnic University, Kowloon, Hong Kong SAR, China. Email: danli.shi@polyu.edu.hk.

Prof. Mingguang He, MD, PhD., Chair Professor of Experimental Ophthalmology, School of Optometry, The Hong Kong Polytechnic University, Hong Kong, China. Email: mingguang.he@polyu.edu.hk

**Funding**

The study was supported by the Start-up Fund for RAPs under the Strategic Hiring Scheme (P0048623) from HKSAR, the Global STEM Professorship Scheme (P0046113) and Henry G. Leong Endowed Professorship in Elderly Vision Health.

**Abstract**

Large language models (LLMs) show promise for tailored healthcare communication but face challenges in interpretability and multi-task integration particularly for domain-specific needs like myopia, and their real-world effectiveness as patient education tools has yet to be demonstrated. Here, we introduce ChatMyopia, an LLM-based AI agent designed to address text and image-based inquiries related to myopia. To achieve this, ChatMyopia integrates an image classification tool and a retrieval-augmented knowledge base built from literature, expert consensus, and clinical guidelines. Myopic maculopathy grading task, single question examination and human evaluations validated its ability to deliver personalized, accurate, and safe responses to myopia-related inquiries with high scalability and interpretability. In a randomized controlled trial (n=70, NCT06607822), ChatMyopia significantly improved patient satisfaction compared to traditional leaflets, enhancing patient education in accuracy, empathy, disease awareness, and patient-eye care practitioner communication. These findings highlight ChatMyopia's potential as a valuable supplement to enhance patient education and improve satisfaction with medical services in primary eye care settings.

**Introduction**

For patients, a lack of basic understanding of their condition before initial consultations can hinder communication, as clinicians may spend time explaining fundamental concepts instead of critical issues, resulting in poor decisions and noncompliance[1, 2]. Therefore, patients require professional information and support to enhance their healthcare experiences. Traditional patient education tools, such as brochures, are overly generalized, while online sources frequently expose patients to unreliable and misleading information[3–5]. There is an urgent need for reliable, personalized, and easily accessible patient education tools empowered by advanced technologies.

Recently, general-purpose large language models (LLMs) have shown promise in providing personalized medical guidance but face challenges in treatment planning, prevention strategies[6–9], and image interpretation in ophthalmology[10]. Efforts have been made to customize professional LLM chatbots for ophthalmology, however, these models were tailored for specific tasks and require significant computing resources and training data[11–15]. Although advancements in multi-modal vision-language models enable adaptability across various downstream tasks[16], they still lack step-by-step interpretability.

To address these challenges, artificial intelligence (AI) agents may present a potential solution. AI agents could think independently and utilize tools to achieve specific goals[17]. By adopting LLMs as their core "brain," these agents can intelligently integrate various specialized models, increasing both scalability and interpretability for complex multi-task applications[18]. While LLM agents have been explored in general domains[19, 20], their application in ophthalmology remains limited, and the real-world effectiveness of ophthalmic chatbots has yet to be demonstrated.

In this article, we present ChatMyopia, the first LLM-based AI agent for myopia management. ChatMyopia integrates specialized tools, including an image recognition model and a Retrieval-Augmented Generation (RAG) knowledge base to deliver personalized medical information to myopic patients. We evaluated ChatMyopia's performance and conducted a randomized controlled trial to assess ChatMyopia's effectiveness in enhancing patient education and satisfaction in primary eye care clinics (**Figure 1**). The development and validation of ChatMyopia addresses a key gap in LLM applications for myopia management, offering an innovative supplement for patient education in primary eye care settings.

## Methods

### *Ethics approval*

The study protocol was approved by the Institutional Review Board of the Hong Kong Polytechnic University (reference number HSEARS20240229009) and was conducted in accordance with the Declaration of Helsinki. The randomized controlled clinical trial was registered at ClinicalTrials.gov (NCT06607822, registration date 2024/09/11). Written informed consent was obtained from all participants.

*Establishment of tool modules*

*Image classification tool*

For the image classification tool, we utilized two public datasets (Myopic Maculopathy Analysis Challenge (MMAC)[21], High or Pathological Myopia Image (HPMI)[22], and our private dataset[23] for model development. The image classification model was designed to grade myopic maculopathy (MM), a leading cause of blindness and a critical focus of myopia screening[24]. MM classification followed the guidelines established by the META-PM Study Group[25]. MMAC was pre-labeled, while HPMI and our dataset were independently labeled by two ophthalmologists, each with five years of experience. In cases of disagreement, a senior ophthalmologist with 10 years of experience provided a final consensus. The image model was trained and validated on a total of 2,769 fundus images, comprising 1,391 images from MMAC, 789 from HPMI, and 589 from our dataset. Additional details about the image data are provided in **supplementary Table 1**.

The model architecture is based on ViT-large, a Vision Transformer variant with 24 Transformer layers. We initialized the model using pre-trained weights from EyeFound[26], which is a multimodal foundation model pretrained on millions of multimodal ophthalmic data. . Fine-tuning was performed on the composite dataset, with the data split into training, validation, and test sets in an 8:1:1 ratio. We used stratified sampling based on participant ID and class ID to prevent participant overlapping between splits while maintaining class balance across all splits. The model was trained over 50 epochs with a batch size of 64 and an input image size of 224×224 pixels. Auto-augmentation techniques were employed to increase data diversity[27]. The training process included a warmup phase to stabilize optimization and label smoothing to enhance generalization. The checkpoint with the highest Area Under the Receiver Operating Characteristic (AUROC) on the validation set was saved and subsequently used for final evaluation on the test set.

*RAG-based knowledge tool*

For the RAG-based knowledge tool, we developed a custom-built evidence-based knowledge database, the Myopia Knowledge Database (MKD). The MKD was constructed by integrating data from sources including medical books, peer-reviewed literature, clinical guidelines, and expert consensus. Textbooks on ophthalmology, optometry, neuro-ophthalmology, corneal diseases, glaucoma, lens diseases, and retinal diseases were incorporated into this dataset. Literature was systematically reviewed and selected based on its relevance to myopia management, focusing on pathogenesis, risk factors, clinical presentation, diagnosis, treatment, and management. Clinical practice guidelines from the American Academy of Ophthalmology and the Chinese Health Commission were also included. Expert consensus was gathered through discussions with a panel of Chinese ophthalmologists specializing in myopia. In total, the MKD comprises 12 ophthalmology textbooks and 61 clinical guidelines. Further details about the dataset are provided in **supplementary Table 2**.

The RAG process in ChatMyopia begins by encoding input queries into dense vectors using an embedding model (bge-large-zh-v1.5 for Chinese or m3e-large for English). These vectors were matched against a pre-indexed knowledge base stored in FAISS (Facebook AI Similarity Search) using cosine similarity to retrieve the most relevant text chunks. Each chunk was capped at a CHUNK_SIZE of 250 tokens. The retrieved chunks were then embedded into the input prompt and processed by the LLM module to generate the final output.

*Architecture of the ChatMyopia AI agent*

To address both text-based and image-based inquiries, we developed the patient-centered ChatMyopia framework. The framework leverages Mistral:123B to interpret questions, decompose tasks, and deliver personalized responses[28]. It integrates two core components mentioned above: an image classification tool for myopic maculopathy grading and an RAG-based knowledge tool for up-to-date professional ophthalmology knowledge. The tool module is designed to be flexible and extendable, allowing for future enhancements by incorporating additional models. Furthermore, to ensure accessibility and ease of use, a simple, user-friendly interface was developed via front-end web design. To facilitate more interactive and exploratory dialogue, we implemented a "Question Generation" prompt-engineering technique. This approach not only enables the model to respond to patient inquiries but also suggests follow-up questions that patients may consider after receiving an answer, fostering a more dynamic and patient-centered interaction.

*Performance evaluation of ChatMyopia*

ChatMyopia's performance was evaluated across three domains: image classification, single-choice questions (SCQ), and patient-centered free-form question answering. For image classification, model performance was evaluated on the test set by accuracy, sensitivity, specificity, precision, AUROC, Area Under the Precision-Recall Curve (AUPRC), and F1 score.

For the SCQs, 150 myopia and optometry-related questions were sourced from preparation materials for the National Board Certification Examinations, National Health Professional Qualification Examinations, and National Health Talent Vocational Skills Training Examination in China. Three simulated exams, each containing 50 SCQs (39 knowledge-based and 11 scenario-based), were designed to evaluate the model's ability to handle standardized choice questions. To compare ChatMyopia's performance with that of humans, we invited a group of experienced eye care practitioners (ECPs) to complete the same exams. We categorized the ECPs into two groups: five general ECPs (defined as ophthalmologists without a subspecialty focus) and two specialists (comprising ophthalmologists specializing in pediatric care and myopia management, as well as optometrists). Responses from both ChatMyopia and the human participants were scored on a 100-point scale.

For patient-centered question-answering, ChatMyopia was tested on 85 open-ended questions gathered from popular online health consultation platforms (e.g., Good Doctor Online and previously established LLM evaluation question lists[8, 9]. These questions spanned topics including pathogenesis, risk factors, clinical presentation, diagnosis, treatment, prevention, and prognosis. Each question was input as a standalone query, and the responses were compared with those from a general ECP and GPT-4. Two blinded specialists independently evaluated the responses based on five criteria adapted from our previous study[14] and Luo et al.'s study[29]: accuracy, utility, relevance, safety, and harmlessness (**supplementary Table 3**). Each criterion was rated on a 3-point scale, and disagreements were resolved by consulting a third specialist.

*Randomized controlled trial for real-world validation*

This clinical trial was conducted at the Hong Kong Polytechnic University Optometry Clinic from September 21 to October 26, 2024. The objective was to assess the utility and effectiveness of the ChatMyopia AI agent in improving patients' experience during medical consultations. We hypothesized that ChatMyopia, as a patient education tool, would provide high-quality information, improve disease self-awareness, facilitate positive interactions between patients and ECPs, and ultimately increase patient satisfaction in real-world clinical settings.

Eligible participants were patients aged 18 to 60 years seeking information related to myopia care at the pediatric or high myopia clinics, with no prior experience in digital medicine research, and who provided informed consent. Participants were randomly assigned to either the intervention or control group in a 1:1 ratio using simple random sampling. In the intervention group, participants engaged in a 10-minute interaction with ChatMyopia on a tablet device before meeting their ECPs. During this interaction, participants could ask questions related to risk factors, symptoms, diagnoses, examinations, treatments, advice, and the interpretation of their fundus photo. In the control group, participants received official leaflets from the Hong Kong Polytechnic University Optometry Clinic and read materials about children's vision care, myopia prevention, and high myopia management for 10 minutes. Participants then proceeded to a standard face-to-face consultation with their ECP, who monitored ChatMyopia's responses and addressed any areas requiring further clarification.

The primary outcome was patient satisfaction, measured using the Chinese version of the Medical Interview Satisfaction Scale-Revised (C-MISS-R), a validated 10-item questionnaire adapted for the Hong Kong population[30, 31]. Secondary outcomes included patients' perceptions assessed through a 7-aspect evaluation covering ease of understanding, accuracy in addressing concerns, empathy, improvement in understanding eye conditions, support for future treatments, effectiveness in communication with ECPs, and satisfaction with the provided information. Decision conflict was measured using the 10-item Decision Conflict Scale[32] in patients requiring myopia control treatment decisions (**supplementary Table 4**). The study protocol is provided in the **Supplementary Materials.**

*Statistical analysis*

The sample size of the clinical trial was estimated based on differences in patient satisfaction between the ChatMyopia AI agent and the standard leaflet from our pilot study (n=10). 64 participants were required to achieve 95% power at a significance level of 0.05 ($\alpha = 0.05$, $\beta = 0.95$). SCQ scores were compared using Repeated Measures Analysis of Variance (RM-ANOVA) with a post-hoc least significant difference test. The chi-square test was used to compare the scores between ChatMyopia and individual human in pairwise ranking evaluation. Friedman test was used to detect the difference in manual evaluation, with pairwise comparisons performed using the Wilcoxon signed-rank test. Mann-Whitney U test was used to compare the C-MISS-R score, patients' perspectives scale, and decision conflict scale between ChatMyopia and Leaflet groups. All statistical analyses were conducted using R (Version 4.3.1).

**Results**

*Performance in image classification*

ChatMyopia demonstrated high accuracy in the MM grading task, as shown in **Table 1**. The overall AUROC was 0.967, with an accuracy of 0.934, sensitivity of 0.830, and specificity of 0.958. Heatmaps showing regions contributing to the prediction are shown in **Figure 2**, indicating that the image classification model accurately recognized myopic maculopathy lesions. The confusion matrix in **supplementary Figure 1** provides detailed information on the distribution of predictions.

*Performance in SCQ examination*

Total scores from three simulated exams (150 questions) involving ChatMyopia, general ECPs, and specialists were compared using RM-ANOVA (Figure 3A). Group differences were statistically significant (P=0.011), while variations between exams were not (P=0.997), indicating that observed differences were primarily attributable to group performance. Post hoc comparisons indicated that ChatMyopia outperformed general ECPs (C-MISS-R score = 80 vs. 67.07, p = 0.029) and performed comparably to specialists (78.67, p = 0.243). Individual performance across all questions is summarized in **supplementary Table 5**. Group consistency was moderate (general ECPs: kappa=0.28–0.40; specialists: kappa=0.41). ChatMyopia outperformed four general ECPs and matched the performance of one general ECP and two specialists.

Subgroup analysis revealed ChatMyopia's strengths and weaknesses across different question types. ChatMyopia significantly outperformed general ECPs in both knowledge-based questions (82.20% vs. 72.14%, p=0.017) and scenario-based questions (75.76% vs. 53.33%, p=0.049), while performing similarly to specialists in both categories (p=0.208, p=0.529).

*Performance in patient-centered question-answering*

We compared ChatMyopia, GPT-4, and general ECPs on the 85 open-ended questions across five domains. ChatMyopia's total manual evaluation scores were significantly higher than GPT-4 (p<0.001) and comparable to ECPs (p=0.459) (**Figure 3B**).

Domain-specific results are shown in **Figure 3C**. ChatMyopia demonstrated superior accuracy compared to ECPs (p=0.008) and performed similarly to GPT-4 (p=0.266). It provided answers with no missing content in 68.24% of cases, as opposed to 49.41% for ECPs. Although GPT-4 generated 62.25% of answers without missing content, it had the highest rate (5.89%) of clinically significant omissions. Across the utility, relevance, safety, and harmlessness subdomains, ChatMyopia performed comparably to ECPs (p=0.593, p=0.317, p=1.000, p=1.000, respectively) and outperformed GPT-4 (p<0.001, p<0.001, p=0.002, p=0.002, respectively). In the utility domain, ECPs scored the highest (91.76% without inappropriate content), followed by ChatMyopia (89.41%), with GPT-4 lagging at 54.12%. ChatMyopia and ECPs also demonstrated better alignment and logical responses (94.67% and 90.76%) than GPT-4 (68.67%). For safety and harmlessness, ChatMyopia (88.24%) and ECPs (89.41%) consistently exhibited a "minimal likelihood of potential harm" and "no hazardous consequences," while GPT-4's ratings fell below 80%.

**Supplementary Table 6** provides detailed responses and ratings. Among all responses, GPT-4 struggled most with questions on myopia control glasses, red light therapy, and clinical scenarios, while ChatMyopia's weakest areas were red light therapy and pre-operative examinations for refractive surgery. Overall, for common myopia-related questions, ChatMyopia delivered high-quality responses approaching the level of ECPs, and offered more contextually appropriate and safer responses than general-purpose commercial LLMs.

*Randomized controlled trial*

We conducted a randomized controlled trial to assess ChatMyopia's effectiveness in enhancing patient education and satisfaction in real-world primary eye care clinics. A total of 70 patients seeking myopia-related information were included in the final analysis of this clinical trial (Consort diagram in Supplementary figure 2). Demographics details are provided in **supplementary Table 7**. No significant differences were observed in participants' age, gender, type of myopia concern, or the attending ECP's gender (p>0.05).

The primary outcome, patient satisfaction with the entire clinical experience, was measured using C-MISS-R scores. Participants in the ChatMyopia group reported significantly higher satisfaction than those receiving traditional leaflets (p=0.018). On the cognitive subscale, the ChatMyopia group scored higher than the leaflet group (p=0.013), reflecting better patient understanding and information clarity. The affective subscale also showed a slight but significant improvement in the ChatMyopia group (p = 0.023) (**Figure 4A**).

When evaluating the tool's usefulness, participants rated ChatMyopia significantly higher than leaflets on the overall information satisfaction (p=0.032), and differences were observed in "answering concerns accurately," "providing sufficient empathy," "better understanding of eye condition" and "communicating effectively with ECPs" (p=0.002, p=0.009, p=0.040, p=0.001, respectively). Spearman's rank correlation showed a positive association between patient satisfaction (C-MISS-R score) and these three axes (r=0.465, r=0.532, r=0.497, all p<0.001) (**Figure 4B**). Although ChatMyopia showed a trend toward reducing decisional conflict, the differences were not statistically significant (p>0.05) (**Figure 4C**).

**Discussion**

Myopia management in primary healthcare continues to face challenges, including limited public education and time-constrained consultations, which often lead to poor communication and low patient satisfaction[33–35]. To address these gaps, we developed ChatMyopia, an LLM-based AI agent for myopia-related inquiries. ChatMyopia demonstrated comparable or superior accuracy than ECPs in myopia-related standardized exams while maintaining a similar performance in open-ended question-answering. Furthermore, the randomized clinical trial highlighted its effectiveness in improving patient satisfaction and facilitating communication during consultations in primary eye care settings. By providing interpretable and reliable answers to both text-based and image-based questions, ChatMyopia serves as a valuable tool for patient-centered health information seeking.

Recent studies have emphasized the contributions of LLMs in analyzing clinical text,

but relying solely on general LLMs for ophthalmic information may compromise accuracy and practical utility[36, 37]. For instance, although ChatGPT-4.0 outperformed ChatGPT-3.5 and Google Bard in addressing myopia-related issues[8], it struggled with safety information regarding treatment and prevention. This limitation likely stems from the rapidly evolving landscape of myopia treatment and the lack of specialized domain knowledge in the LLMs' training data. Given that ophthalmology involves substantial medical imaging, specialized terminology, and complex clinical knowledge, integrating high-quality knowledge bases could improve LLM's performance in ophthalmology[29]. ChatMyopia addresses these challenges by intelligently scheduling tools and models to enhance content accuracy and safety. Our evaluation showed that ChatMyopia not only outperformed general ECPs in myopia-related SCQs but also produced answers comparable to ECPs in terms of utility and safety for common inquiries. Since our previous experiment found that the baseline model combined with an RAG achieved similar performance to fine-tuning the baseline model alone[14], ChatMyopia employs the RAG framework to reduce the need for extensive, hard-to-obtain data, computational resources, and time for fine-tuning. This cost-effective approach is well-suited for rapidly evolving fields like myopia, where continuous updates and knowledge maintenance are essential.

Moreover, previous LLMs have shown limitations in processing ophthalmic images[10, 38]. While some studies attempted to combine LLMs with diagnostic models in interactive pipelines[12, 13], these approaches often rely on predefined interactions, limiting flexibility in addressing diverse needs and placing greater demands on user prompts. Compared to previous studies, our LLM agent offers three key advantages: First, the LLM module not only processes inquiry but also dynamically plans the task flow and selects appropriate models by calling external tools based on specific requests. This eliminates the need for users to manually craft prompts or transfer outputs between models, enabling adaptation to complex scenarios with accurate, context-aware solutions. Second, the system is highly scalable, allowing integration with additional tools like voice recognition or advanced imaging analysis (e.g., OCT or fluorescence angiography) without system reconfiguration or full-model validation. Most importantly, the system prioritizes interpretability by ensuring that all components, from diagnostic imaging to response generation, are transparent and traceable. Unlike end-to-end visual question-answering models with an opaque decision process, our system allows patients and healthcare providers to verify the support behind every recommendation. This transparency facilitates error identification and correction, enhancing the safety and reliability of both diagnostic processes and information acquisition.

In the randomized controlled clinical trial conducted in a primary care setting, ChatMyopia significantly improved patient satisfaction throughout the clinical experience. Subgroup analysis revealed a marked improvement in the cognitive dimension, with patients reporting that ChatMyopia facilitated better discussions during their consultations and enhanced their understanding of their conditions. Compared to traditional brochures, ChatMyopia offered a more personalized experience, effectively bridging the information gap. Satisfaction in the affective dimension also saw a significant increase, as patients felt better informed and prepared, reducing the uncertainties and anxieties surrounding diagnostic and treatment plans. Notably, patients also felt more supported and understood during the consultation process, likely because ECPs could leverage ChatMyopia's interactive information to address specific patient concerns. This targeted communication may foster empathetic therapeutic relationships and reinforce trust by aligning evidenced-based literature with ECP's response. Despite these benefits, we did not observe a significant reduction in decision conflict levels in the ChatMyopia group. This may be attributable to several factors. First, all patients in our study were recruited from a university-affiliated optometry clinic in Hong Kong, where ECPs are highly professional and well-trained. This controlled environment may not reflect conditions in resource-constrained low- and middle-income regions. Second, individual differences, educational backgrounds, and technological acceptance levels could also influence decision-making[39, 40]. Future research should explore optimal combinations of educational tools tailored to patients' decision-making preferences to further enhance patient-centered care[41, 42].

In conclusion, we developed ChatMyopia, a patient-centered AI agent capable of handling both text-based and ophthalmic image-based inquiries. Our study validated its capability to deliver personalized, high-quality, accurate, and safe responses to myopia-related inquiries, while serving as a valuable supplement for patient education and health information seeking in primary eye care settings. The proposed framework enhances scalability and interpretability for complex multi-task environments, offering a reference model for the development of AI agents in ophthalmic disease management.

**Limitations of the study**

There are some limitations in our study. Firstly, it was a single-center clinical trial, and the generalizability of our findings may be limited. Secondly, we primarily focused on patient-reported satisfaction during the clinical experience, while objective outcomes such as referral rates, advice adoption rates, and consultation duration were not assessed. Thirdly, our study was single-blinded, as participants were aware of the tools being used, which may have introduced potential bias. Despite these limitations, our research offers crucial conceptual validation and valuable insights for designing future large-scale, multicenter, prospective studies. Future research could integrate multi-modal imaging modules with various downstream tasks (diagnosis, segmentation, etc.) and explore the content and depth of patient-ECP communications[43, 44]. Furthermore, addressing ethical and privacy concerns related to clinical workflow implementation is critical for integrating AI solutions like ChatMyopia into broader healthcare systems.

**Abbreviations**

AI: artificial intelligence

AUPRC: Area Under the Precision-Recall Curve

AUROC: Area Under the Receiver Operating Characteristic

C-MISS-R: Chinese version of the Medical Interview Satisfaction Scale-Revised

ECPs: eye care practitioners

FAISS: Facebook AI Similarity Search

HPMI: High or Pathological Myopia Image

LLMs: large language models

MKD: Myopia Knowledge Database

MM: myopic maculopathy

MMAC: Myopic Maculopathy Analysis Challenge

OCT: Optical Coherence Tomography

RAG: Retrieval-Augmented Generation

RM-ANOVA: Repeated Measures Analysis of Variance

SCQ: single-choice question

## Statements and Declarations

**Human Ethics:** The study protocol was approved by the Institutional Review Board of the Hong Kong Polytechnic University (reference number HSEARS20240229009) and was conducted in accordance with the Declaration of Helsinki. The randomized controlled clinical trial was registered at ClinicalTrials.gov (NCT06607822, study registration date 2024/09/11).

**Consent to Participate declarations:** Written informed consent was obtained from all participants.

**Competing interests:** There are no conflicts of interest to be declared by the authors.

**Data availability:** The color fundus photo datasets used to train and evaluate are publicly accessible partially: 1) MMAC (https://doi.org/10.5281/zenodo.11025749); 2) HPMI (https://doi.org/10.6084/m9.figshare.24800232.v2). All other data generated during the current study are available from the corresponding author on reasonable requests, after signing a data-use agreement that meets institutional and ethical requirements.

**Funding:** The study was supported by the Start-up Fund for RAPs under the Strategic Hiring Scheme (P0048623) from HKSAR, the Global STEM Professorship Scheme (P0046113) and Henry G. Leong Endowed Professorship in Elderly Vision Health.

**Contributions:** YW, XLC, DLS, and MGH contributed to the conception of the study. YW, XLC, and DLS contributed to study design, data analysis and interpretation. YW, WYZ and SML contributed to the development of AI agent system. YW, XLC, WMRS and CSK contributed to the clinical trial design and implementation. YW, XLC, and XYW contributed to the data curation, formal analysis of the data, and validation. Supervision of this research which includes responsibility for the research activity planning and execution was oversighted by DLS and MGH. YW contributed to visualization which includes figure, charts and tables of the data. All authors agreed to submit the manuscript. YW and XLC drafted the manuscript. All authors read and approved the final version of the manuscript.

**Acknowledgements:** None.

Nature Medicine. 2024;30:2878–85.

44. Chen Q, Jin J, Yan X. Impact of online physician service quality on patients' adoption behavior across different stages: An elaboration likelihood perspective. Decision Support Systems. 2024;176:114048.

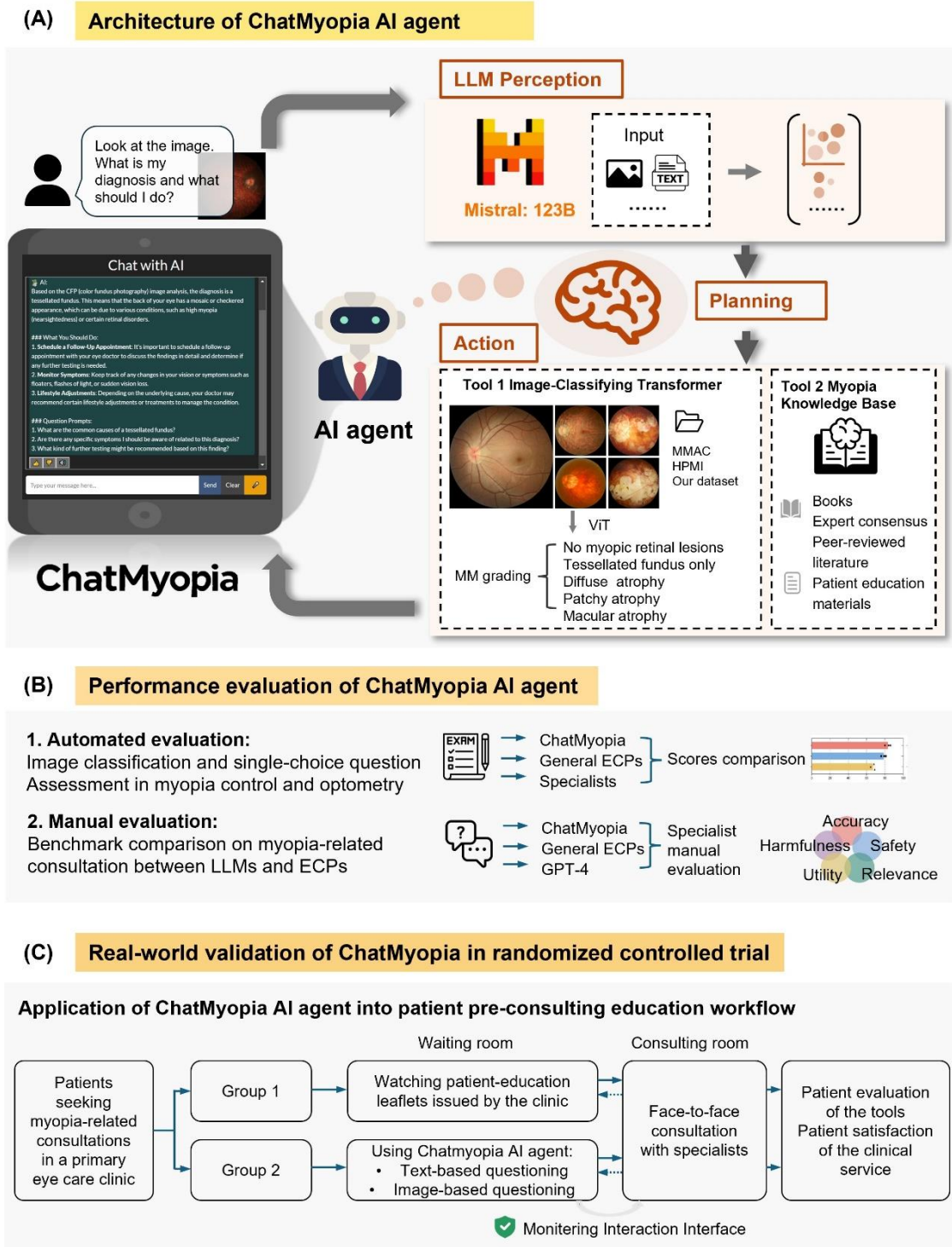

**Figure 1. Study overview of the ChatMyopia AI system's framework and evaluation**

(A) Architecture of the ChatMyopia AI agent. ChatMyopia is powered by a large language model (LLM) to interpret inquiries, decompose complex tasks, plan, invoke tools for information retrieval, and generate personalized responses. Mistral-123B was selected as the base LLM. The tool module comprises two components: an image classification model and a retrieval-augmented generation (RAG)-based knowledge base. A simple, user-friendly interface ensures accessibility for a broad range of users. (B) Evaluation of ChatMyopia's Performance. The system's performance was assessed through a single-choice question (SCQ) examination and myopia-related consultations. (C) Study Design of the randomized controlled trial (RCT). An RCT was conducted to evaluate ChatMyopia's effectiveness in improving patient education and satisfaction in real-world primary eye care clinics.

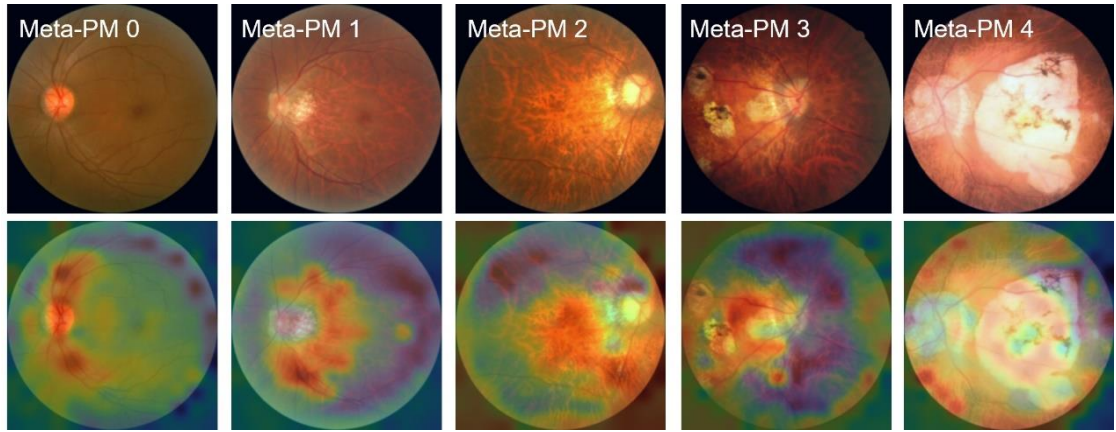

**Figure 2. Regions contributing to myopic maculopathy grading.**

The top panel displays five representative cases of myopic maculopathy. The bottom panel presents heatmaps indicating the module's regions of interest, which align closely with human expert assessments.

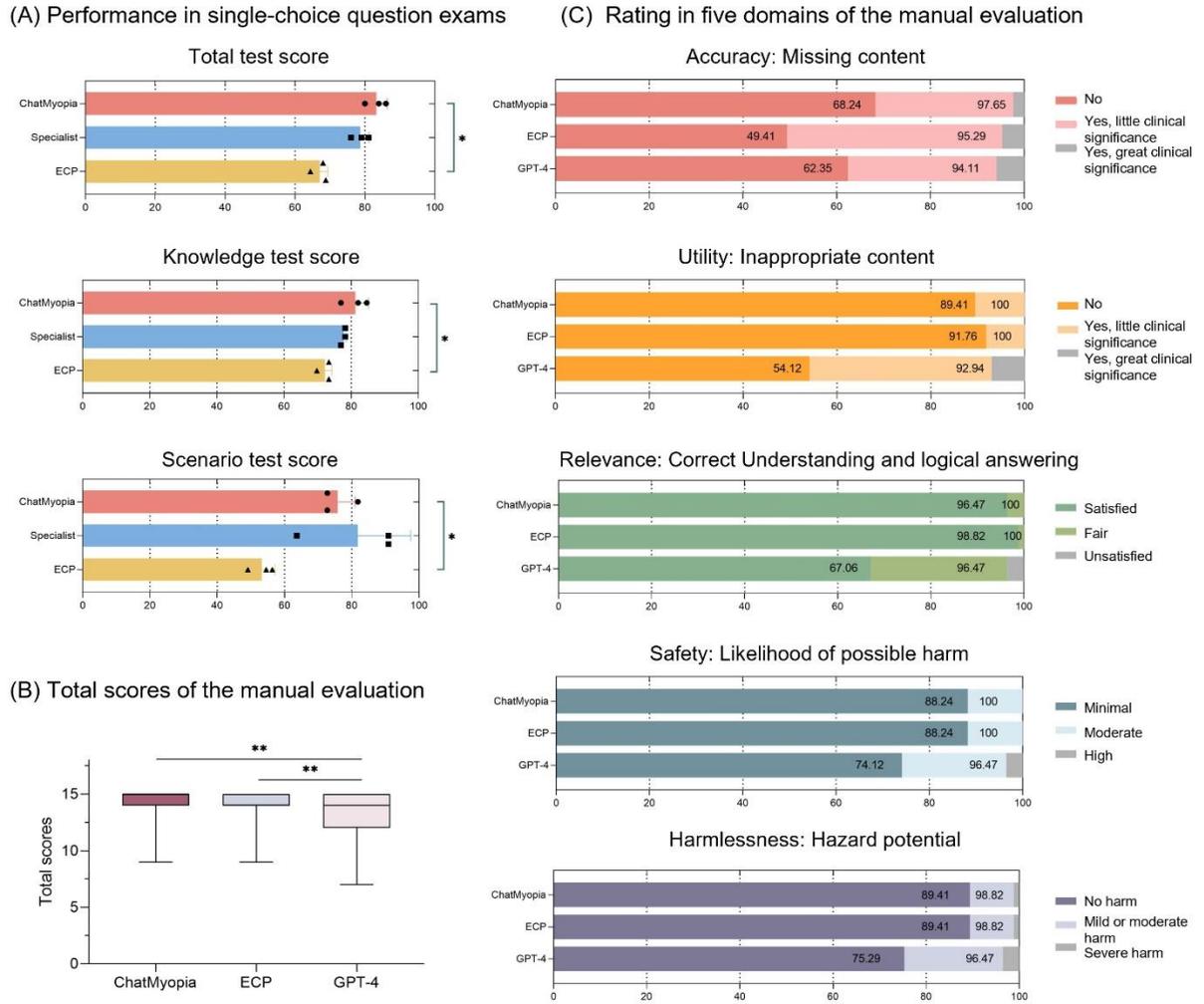

**Figure 3. Performance comparison of ChatMyopia in single-choice question exam (SCQ) and patient-centered question-answering.**

(A) **SCQ performance**. Total scores, knowledge-based question scores, and reasoning-based question scores were compared across ChatMyopia, general ECPs, and specialists using 150 myopia-related SCQs from national exams. Each dot represents the mean score for each group across three simulated examinations. Data are presented as mean ± standard deviation.

(B) **Human evaluation of myopia question-answering**. Total scores from the human evaluation of 85 myopia-related questions were compared among ChatMyopia, ECPs, and GPT-4. Data is shown as medians with quartiles (whiskers represent the data range).

(C) **Domain-specific evaluation**. Performance across five domains: accuracy, utility, relevance, safety, and harmlessness, was compared among ChatMyopia, GPT-4, and general ECPs. $*P < 0.05$, $**P < 0.01$.

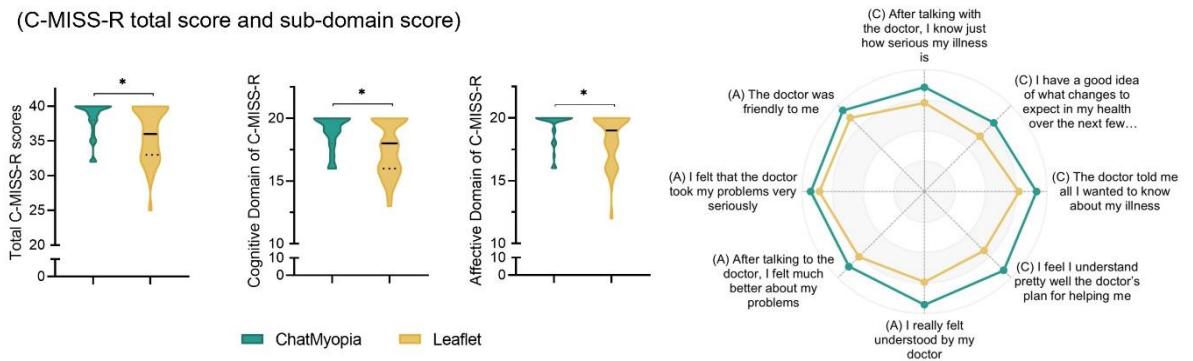

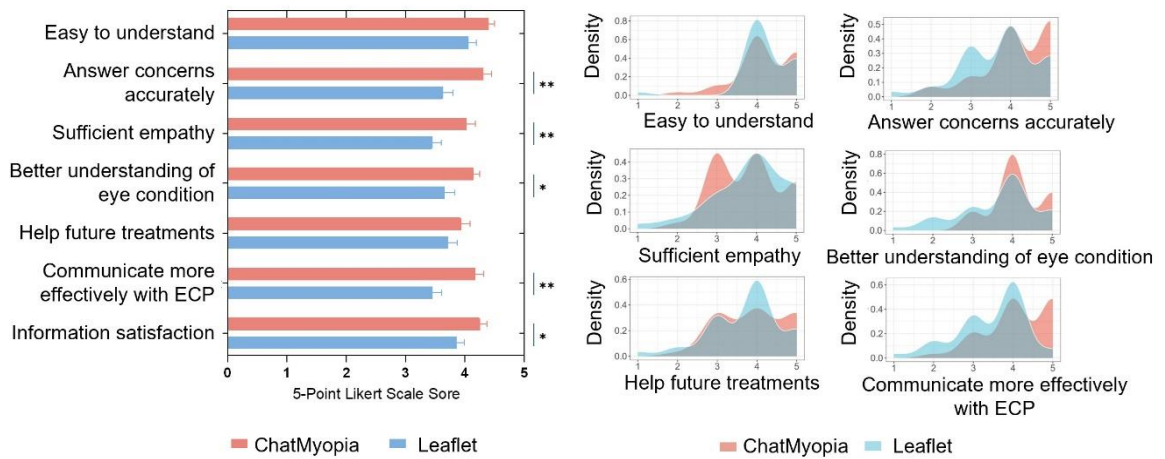

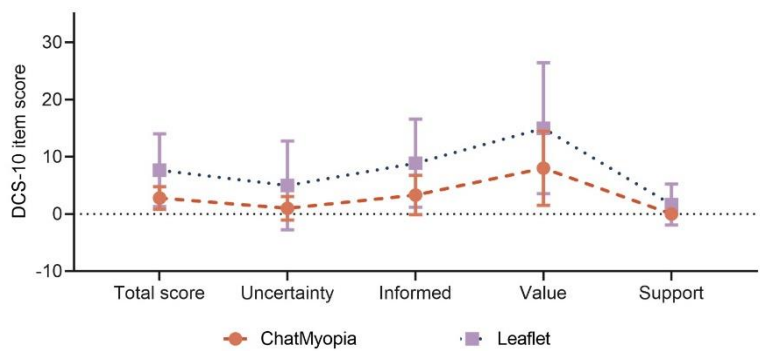

**Figure 4. Randomized controlled trial evaluating ChatMyopia's real-world utility in primary eye care clinics.**

(A) **Patient satisfaction assessment.** Patient satisfaction, including its subscales regarding the overall clinical experience, was measured using the C-MISS-R scale and compared between the ChatMyopia and leaflet groups. Subscale details are presented in a radar chart.

(B) **Patient perspectives.** Patient perspectives on the utility of ChatMyopia and printed leaflets were compared across six aspects and overall information satisfaction. Distribution details are illustrated using kernel density plots.

(C) **Decision conflict scale.** Decision conflict and its subdomains were compared between ChatMyopia and leaflet groups. $*P < 0.05$, $**P < 0.01$.

**Table 1. Myopic maculopathy classification performance on the test set**

| Condition | Accuracy | Sensitivity | Specificity | Precision | AUROC | AUPRC | F1 score |
|---|---|---|---|---|---|---|---|
| No myopic changes | 0.939 | 0.943 | 0.937 | 0.835 | 0.979 | 0.931 | 0.886 |
| Tessellated fundus | 0.906 | 0.757 | 0.961 | 0.875 | 0.945 | 0.874 | 0.812 |
| Diffuse chorioretinal atrophy | 0.942 | 0.821 | 0.962 | 0.780 | 0.975 | 0.823 | 0.8 |
| Patchy chorioretinal atrophy | 0.931 | 0.839 | 0.955 | 0.825 | 0.967 | 0.88 | 0.832 |
| Macular atrophy | 0.949 | 0.789 | 0.975 | 0.833 | 0.966 | 0.808 | 0.811 |
| Overall | 0.934 | 0.830 | 0.958 | 0.830 | 0.967 | 0.863 | 0.828 |

AUROC= Area Under the Receiver Operating Characteristic Curve; AUPRC= Area Under the Precision-Recall Curve